\documentclass[a4paper]{article}
\usepackage[bookmarks=true]{hyperref}
\hypersetup{
colorlinks=true,
linkcolor=red,
citecolor=blue,
filecolor=black,
urlcolor=blue,
}
\usepackage[english]{babel}
\usepackage[utf8]{inputenc}
\usepackage{amsmath}
\usepackage{mathtools}
\usepackage{graphicx}
\usepackage[colorinlistoftodos]{todonotes}
\usepackage{textcomp}

\usepackage{geometry}
\newgeometry{top=1.0in, bottom=1.0in, left=1.0in, right=1.0in}
\usepackage{indentfirst}
\usepackage{float} 
\usepackage{caption}
\usepackage{subcaption}
\numberwithin{equation}{section}
\usepackage{enumitem}
\usepackage{wrapfig}
\usepackage[version=3]{mhchem}
\usepackage{pdfpages}

\usepackage{fixltx2e}
\usepackage{etoolbox}
\usepackage{cancel}
\usepackage{epstopdf}

\usepackage{authblk}

\usepackage{caption}
\usepackage{subcaption}

\usepackage{array}
\newcolumntype{L}[1]{>{\raggedright\let\newline\\\arraybackslash\hspace{0pt}}m{#1}}
\newcolumntype{C}[1]{>{\centering\let\newline\\\arraybackslash\hspace{0pt}}m{#1}}
\newcolumntype{R}[1]{>{\raggedleft\let\newline\\\arraybackslash\hspace{0pt}}m{#1}}

\newcommand{\parenthesis}[1]{\left( #1 \right)}

\newcommand{\absbracket}[1]{\left\lvert #1 \right\rvert}

\title{Feasibility Study of a Laser-Based Approach for Diagnosing Deuterium Neutrals in the Edge of Fusion Devices}
\author[1]{David Feng}
\author[2]{Ahmed Diallo}
\author[1,2]{Mikhail N. Shneider}

\affil[1]{Department of Mechanical \& Aerospace Engineering, Princeton University, Princeton, NJ 08544}
\affil[2]{Princeton Plasma Physics Laboratory, Princeton, NJ 08540}

\date{\today}

\begin{document}

\maketitle

\section{Motivation and Goal}

In magnetically-confined plasmas of tokamaks, neutral deuterium/hydrogen (D/H) atoms play a role in energy, momentum, and particle balance, as well as the stabilization of plasma turbulence. One key important fusion performance parameter is the pedestal density. Understanding the pedestal density formation is critical for the development of predictive model of future fusion devices. Typically, measurements of the neutrals are obtained using optical emission spectroscopy of the Lyman alpha lines, which is a line-integrated measurement. The plasma in tokamaks is characterized by a high density of electrons and ions and a relatively low concentration of neutral hydrogen atoms, which could make direct measurement of density seemingly impossible at first. \\

We propose a laser-based method that allows for accurate measurement of both the spatial and absolute magnitude of the neutral D/H with minimal knowledge of the radial profiles of electron temperatures and densities. This relies on the fact that the neutral spectral profile can have a larger peak than the electron spectral profile and thus make the neutral density signal resolvable. In practice, this method can be co-located with Thomson scattering systems and is referred to as \textit{laser Rayleigh scattering} (LRS). \\

More specifically, we assess and evaluate the LRS method for two test cases: in the midplane radii of the National Spherical Torus Experiment (NSTX), and in the small angle slot divertor configuration of DIII-D. Preliminary simulations and calculations will determine the feasibility of LRS in the presence of incoherent Thomson scattering under neutral densities ranging from $ 10^{13} $ to $ 10^{21} \ \text{m}^{-3} $. Wavelength dependence of LRS will be evaluated to determine the boost in the signal and photon generation capability. \\


\section{The Rayleigh Scattering Diagnostic: A Brief Overview}

LRS is the elastic scattering of light by neutral atoms and molecules much smaller than the wavelength of the incident laser light. In the classical sense, it is the electromagnetic radiation emitted by an oscillating dipole moment (the atom/molecule) that is induced by an incoming electric field (the laser):
\begin{align}
    \textbf{p} = \alpha \ \textbf{E}(\textbf{r},t) \label{p}
\end{align}

The polarizability $ \alpha $ characterizes the electric response specific to an individual class of scatterers, such as atoms, molecules, and even ions. It is a tensor quantity that can be simplified and reduced depending on the shape and orientation of the molecule. If the molecule is far away from resonance, such that the $ \alpha $ is not affected by the frequency of the external electric field, it is called the static polarizability.

\begin{align}
    \alpha_\text{mean}^2 = \frac{1}{9}( \alpha_{xx} + \alpha_{yy} + \alpha_{zz} )^2    
\end{align}

For linear diatomic molecules, the square of effective polarizability averaged over all molecular orientations is \cite{miles_laser_2001}:

\begin{align}
    \alpha_\text{mean}^2 = \frac{45 \alpha^2 + 10 \gamma^2}{45} 
\end{align}
where 
\begin{align}
    \alpha^2 = \frac{1}{9}( \alpha_{xx} + 2\alpha_{yy})^2
\end{align}
\begin{align}
    \gamma^2 = ( \alpha_{xx} - \alpha_{yy} )^2 
\end{align}

X is the axis of symmetry of the molecule, Y and Z are normal to it, and the polarizabilities in the Y and Z directions are equal. \\

If the scatterer is stationary, the emitted radiation is phase-locked to the incident light and the result is a coherent sum of the field plus all other nearby fields. The radiation cancels in all but the forward direction. However, molecular motion in the gas leads to microscopic fluctuations in the gas density that randomize the phase of the emitted radiation and causes it to be incoherent in all but the forward direction. Away from the forward direction, the rapidly changing interference averages out to become proportional to the number of scatterers in the gas (i.e. the number density). The average power of this emitted radiation (i.e. the total scattered average power) is given by:
\begin{align}
    P = n V I_\text{L} \sigma_{\text{RS}} \label{P_RS}
\end{align}
where $ N $ is the number density of the gas, $ V $ is the volume of scatterers, $ I_\text{L} $ is the initial laser intensity, and $ \sigma_{\text{RS}} $ is the Rayleigh scattering (RS) cross-section of a single scatterer (or species). A more thorough description of LRS can be found in Miles et al. \cite{miles_laser_2001}. When LRS is used for diagnostics (e.g. single or multiple photodetectors), it is more practical to cast Equation \ref{P_RS} in differential form per unit solid angle $ \Omega $:
\begin{align}
    d_{\Omega} P = n V I_\text{L} d_{\Omega} \sigma_{\text{RS}} \label{dP_RS}
\end{align}
where $ d_{\Omega} \sigma_{\text{RS}} $ is the differential RS cross-section. It can be related to $ \alpha $ via:
\begin{align}
    d_{\Omega} \sigma_{\text{RS}} = \alpha_{\text{V}}^2 k_{\text{L}}^4 \sin^2{\phi} \label{DiffCrossSec}
\end{align}
where $ \alpha_{\text{V}} = \alpha / 4 \pi \varepsilon_0 $, $ k_{\text{L}} $ is the laser wavevector, and $ \phi $ is the azimuthal angle of the dipole radiation pattern. The total RS power is therefore:
\begin{align}
    P_{\text{RS}} = \eta n V I_{\text{L}} \int_{\Delta \Omega}  d_{\Omega} \sigma_{\text{RS}} \ d_{\Omega}
\end{align}
after multiplying by the appropriate optical, collection, and detector efficiencies $ \eta $. It is evident that LRS can be used to determine aggregate number densities of the gas-phase system, including those of near-wall tokamak devices. Note that for multiple species present within the gas-phase system, $ \sigma_{\text{RS}} $ becomes the weighted cross-section by mole fraction $ X_i $:
	
\begin{align}
    \sigma_{\text{RS}} = \sum_i X_i \sigma_i
\end{align}
	
For species of interest in nuclear fusion devices (i.e. typically species generated in the outer midplane regions of tokamaks), the table of polarizabilities is given by:

\begin{table}[H]
\centering
\begin{tabular}{ |l|l|l|L{5cm}| }
\hline
Species & Mean polarizability ($A^3$) & Ref. & Note \\ \hline
\ce{H}       & 0.6668 & \cite{sadeghpour_rayleigh_1992} &  Calculation compared with previous data in close agreement. \\ \hline
\ce{H_2}      &  0.7894 & \cite{milenko_temperature_1972} & Calculation for the rotational quantum no. $ j = 0 $ in the electronic ground state. \\ \hline
\ce{H_2+}     &  0.4696 & \cite{hilico_polarizabilities_2001}  &  Calculation for the \ce{^1S^e} $ v = 0 $ state. \\ \hline
\ce{D}       &  0.7587 & CCCDB & Calculation via DFT and the basis set aug-cc-pVTZ. No other data exists in the literature as far as the authors know. \\ \hline
\ce{D_2}      &  0.7954 & \cite{milenko_temperature_1972} & Calculation for the rotational quantum no. $ j = 0 $ in the electronic ground state. \\ \hline
\ce{D_2+}     &  0.4552 & \cite{hilico_polarizabilities_2001} & Calculation for the \ce{^1S^e} $ v = 0 $ state. \\ \hline
\end{tabular}
\caption{Static mean polarizabilities for the species considered in the preliminary study, along with their references. CCCDB stands for the computational chemistry comparison and benchmark database operated by the National Institute for Standards and Technology (NIST).}
\end{table}

One major disadvantage with LRS is that background (BG) scattering from windows, surfaces, and Mie scattering can override the LRS signal due to typically being orders of magnitude stronger. The Mie and dust scattering is random and can be removed via post-processing, but the window and surface reflections will prevent the LRS from being visible, and if the incident laser beam or sheet is too close to a surface, then the desired image may not be properly recorded. However, since the BG scattering occurs at the laser wavelength and is very narrow in linewidth, it can be properly attenuated using an iodine filter, which has absorption bands at 532 nm. The combination of a very narrow linewidth laser and an atomic/molecular notch filter is known as filtered Rayleigh scattering (FRS), and is described in Miles et al. \cite{miles_flow_2001}. \\

Another disadvantage particular to plasma environments is the presence of Thomson scattering (TS). If the laser wavelength is shorter than the Debye length of the plasma, the scattering from electrons is randomly distributed and the scattering is said to be incoherent. For most plasma conditions and measurement geometries, the electron scattering is incoherent. The TS cross-section is typically a few orders of magnitude greater than the RS cross-section for atmospheric and hydrogenic species. However, the TS spectrum is also extremely broad compared to the RS spectrum, often known as the Rayleigh-Brillouin scattering (RBS) spectrum. The TS spectrum is often a few nanometers wide, whereas the RBS spectrum is only a few to tens of GHz wide. Peak values of the signal close to the laser frequency produced by the neutral species can therefore be resolved and detected, and will be the focus of the next section. \\

Electrons and neutrals co-exist in the same environment inside tokamaks, so the LRS will also be coupled to the laser Thomson scattering (LTS). The total scattering cross-section of the electron is given exactly by:
\begin{align}
    \sigma_{\text{TS}} = \frac{8 \pi}{3} \parenthesis{ \frac{e^2}{4 \pi \varepsilon_0 m c^2} }^2
\end{align}

which leads to a value approximately equal to $ 6.65 \times 10^{-29} \ \text{m}^2 $. In comparison, hydrogen has a total RS cross-section of about $ 7.25 \times 10^{-32} \ \text{m}^2 $ at a wavelength of 532 nm. The total cross-section for electrons is three orders of magnitude stronger than that of H, and via Eqn. \ref{P_RS} it would initially seem that four orders of magnitude more number density would be required for the RS of H to become comparable at the very least, e.g. an electron number density of $ 10^{15} \ \text{m}^{-3} $ gives a H number density of $ 10^{18} \ \text{m}^{-3} $. \\

To overcome this fundamental limitation, consider the scattered spectra. Near the laser frequency, at a given electron temperature the peak of the RS spectrum can be equal to or higher for comparable electron and neutral number densities. In the following study, we assume that the RS scattering remains in the kinetic collisionless regime, determined through the y parameter:
\begin{align}
    y = \frac{1}{| k_s | \lambda_\text{MFP} } = \frac{n \sigma_\text{c} }{| k_s |} \propto \frac{P \cdot \lambda_\text{L}}{T}
\end{align}
where $ \nu $ is the laser frequency and $ |k_s| = 4 \pi \sin(\theta/2) / \lambda_\text{L} $ is the scattering wavevector. When $ y \ll 1 $, the RS is in the collisionless kinetic regime, i.e. the Boltzmann equation with no collisional term is used to describe the gas density perturbation, which is approximately Gaussian in nature. Therefore we model the neutral species’ spectra as a Gaussian lineshape:
\begin{align}
    g_\text{RS} (\nu) = \frac{| k_s | / 2\pi}{\Delta v_\text{th}} \cdot e^{(\nu - \nu_0)^2 / \Delta v_\text{th}^2}
\end{align}
where $ \Delta \nu_\text{th} $ is the thermal velocity:
\begin{align}
    \Delta v_\text{th} = \sqrt{ \frac{2 k_B T}{\mu_\text{mol}} }
\end{align}

Note that the lineshape function is normalized such that $ \int g_\text{RS} (\nu) \ d \nu = 1 $. The electron or TS lineshape is modeled using the Salpeter approximation, where the plasma spectrum takes a relatively simple form with both an ion and electron component:
\begin{align}
    g_\text{TS} (\omega) = \frac{2\pi}{|k_s|} \absbracket{ 1 - \frac{\chi_e}{\varepsilon} }^2 f_e\parenthesis{ \frac{\omega}{k_s} } + \frac{2\pi Z}{|k_s|} \absbracket{ \frac{\chi_e}{\varepsilon} }^2 f_i\parenthesis{ \frac{\omega}{k_s} } \label{g_sal}
\end{align}

where $ \omega $ is the angular frequency, $ f_e $, $ f_i $ are the normalized one-dimensional electron and ion velocity distribution functions, $ \chi_e $, $ \chi_i $ are the electron and ion susceptibilities, $ \varepsilon $ is the dielectric function, and $ Z $ is the charge of the ion. Typically, the ion component may be neglected. The approximation assumes a Maxwellian distribution function for the thermal velocities of both the electron and the ion. Through $ f_{e,i} $ and $ \chi_{e,i} $ it is also slightly dependent on the plasma scattering parameter $ \alpha $:

\begin{align}
    \alpha = \frac{1}{| k_s | \lambda_\text{Deb} } \label{scatpar}
\end{align}

where $ \lambda_\text{Deb} = \sqrt{\varepsilon k_B T_e / (n_e e^2) }  $ is the Debye length of the plasma. For $ \alpha \ll 1 $, we have incoherent scattering and the lineshape reflects the distributions of the electrons according to Eqn. \ref{g_sal}. \\


\section{Preliminary Studies: Midplane NSTX}

Consider the inferred deuterium atom and molecule midplane density profiles for the National Spherical Torus Experiment (NSTX) in Figure 2 \cite{stotler_midplane_2015}. The region between $ R = 1.4 $-1.5 m marks the transition region where the molecular densities drop off sharply but the atomic density tails off slower as a result of dissociation due to collisions with ions and electrons.

\begin{figure}[H]
	\centering
	\includegraphics[scale=0.6]{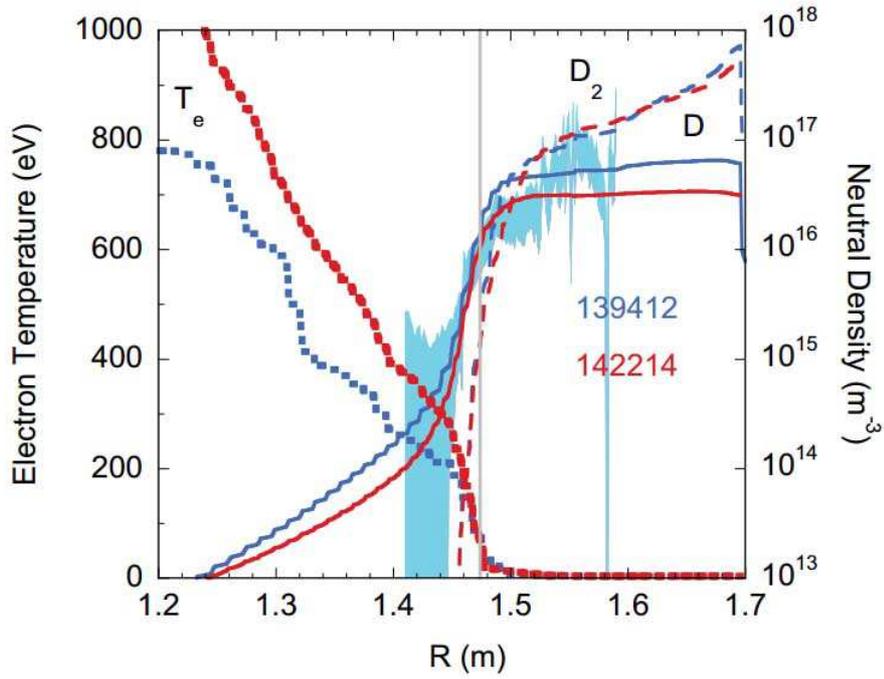}
	\caption{Midplane density profiles for atomic and molecular deuterium for NSTX shots 139412 and 142214. Electron temperature profiles are also provided for reference. Figure is obtained from \cite{stotler_midplane_2015}.}
	\label{stotler}
\end{figure}

To determine the capability of distinguishing LRS from LTS, relevant neutral densities in this transition region were modeled for various electron densities near the laser frequency, i.e. near the peak of the RS spectra. All the following spectra and data in this section will be for the laser wavelength $ \lambda = 532 $ nm. Consider the midplane slice at $ R = 1.4 $ m. Although computational data, derived from ab initio methods, for the polarizability of atomic deuterium (D) exists, it is not reliable and uncorroborated with experimental data or other literature sources; furthermore the same database offers two values which are highly disparate in value. Therefore H will be used in lieu of D.

\begin{figure}[H]
	\centering
	\includegraphics[scale=0.55]{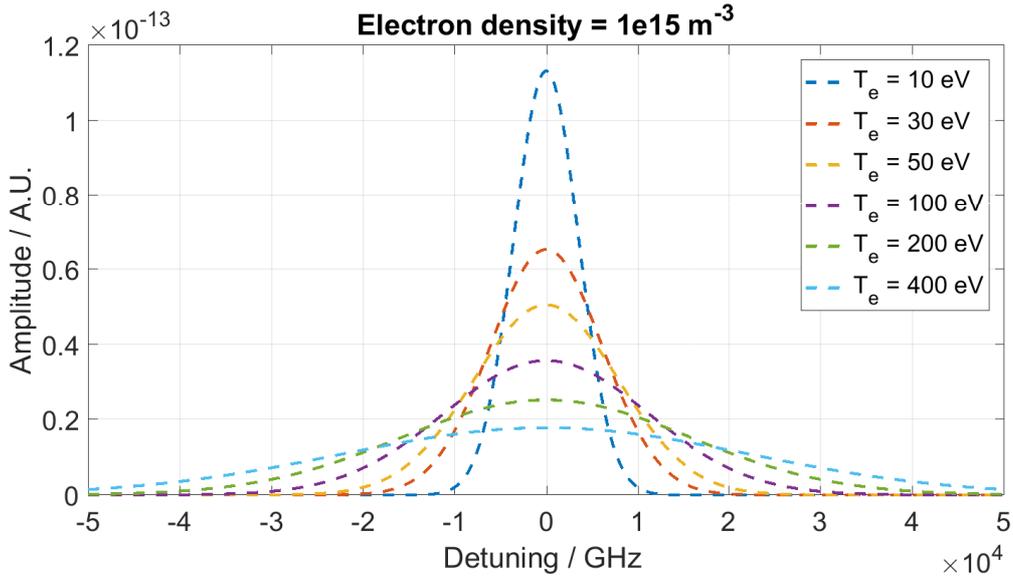}
	\caption{TS spectra at $ n_e = 1 \times 10^{15} \  \text{m}^{-3} $ for multiple electron temperatures. The integrated spectra is normalized to unity.}
	\label{spectra_electrontemp}
\end{figure}

Figure \ref{spectra_electrontemp} demonstrates how the TS spectrum changes with electron temperature. As the temperature rises, the spectrum broadens out because the Maxwellian distributions of the electron (assuming thermodynamic equilibrium) have an inverse proportionality to the thermal speed:
\begin{align}
    f = \parenthesis{ \frac{1}{\pi v_\text{th}^2} }^{1/2} \exp{(-v^2/v_\text{th}^2)}
\end{align}
where
\begin{align}
    v_\text{th} = \sqrt{ \frac{2 k_B T_e }{m_e} }
\end{align}

Therefore as the electron temperature $ T_e $ grows, the distribution function flattens out because of the negative exponential dependence. More importantly, the peak of the TS spectral distribution also drops in order to preserve unity of the spectral integration, which is crucial to observing and resolving the LRS as follows.

\begin{figure}[H]
	\centering
	\includegraphics[scale=0.4]{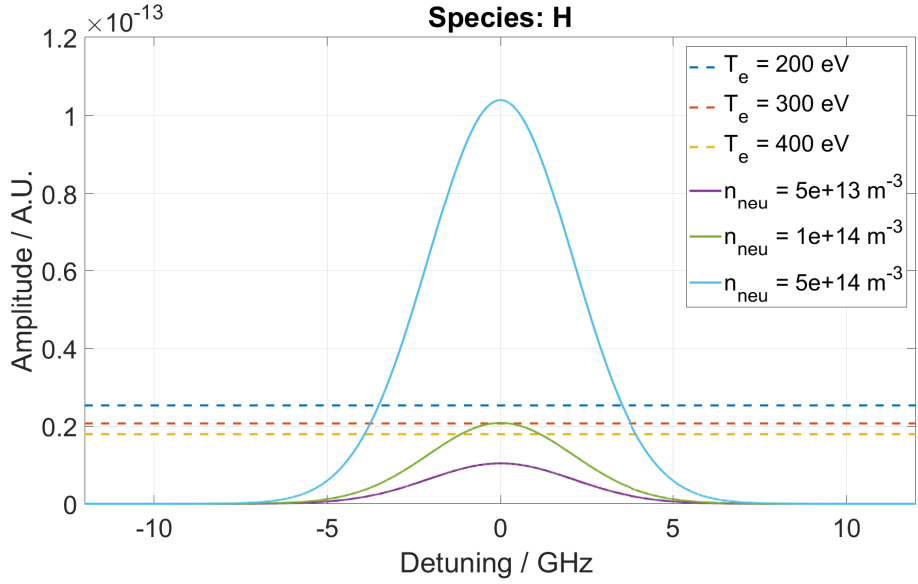}
	\caption{Multiple RS spectra (solid lines) in the presence of TS plasma backgrounds (dashed lines) at $ n_e = 1 \times 10^{15} \  \text{m}^{-3} $ for multiple electron temperatures for H. The peak at $ n_\text{neu} = 5 \times 10^{14} \  \text{m}^{-3} $ is greater than the TS background, which can be detected if the TS is filtered out (see Figure \ref{BPfilter}).}
	\label{spectra_hydrogen}
\end{figure}

Figure \ref{spectra_hydrogen} shows RS spectra for H for up to $ n_\text{neu} = 5 \times 10^{14} \text{m}^{-3} $ in the presence of TS background plasma with electron density $ n_e = 1 \times 10^{15}  \text{m}^{-3} $ and electron temperature $ T_e = 200-400 $ eV. All spectra are scaled with respect to $ n_e $:

\begin{align}
    g_\text{sc} (\nu) = \frac{I_\text{neu}}{I_\text{BG}} g_\text{RS} (\nu - \nu_0; T; \theta) \label{g_sc}
\end{align}

where $ I_\text{neu} / I_\text{BG} $ is the scaling factor for the RS spectra. As $ T_e $ increases, the BG decreases because the TS spectrum is getting broader, and thus the peak must correspondingly flatten if the lineshape is to remain normalized. Note that the TS spectrum is significantly much broader than the RS spectrum, nm compared to pm, respectively. At $ n_\text{neu} = 5 \times 10^{14} \text{m}^{-3} $, the peak is considerably above the BG, and it is clear that values below $ n_\text{neu} = 5 \times 10^{14} \text{m}^{-3} $ will be buried beneath it. \\

\begin{figure}[H]
	\centering
	\includegraphics[scale=0.35]{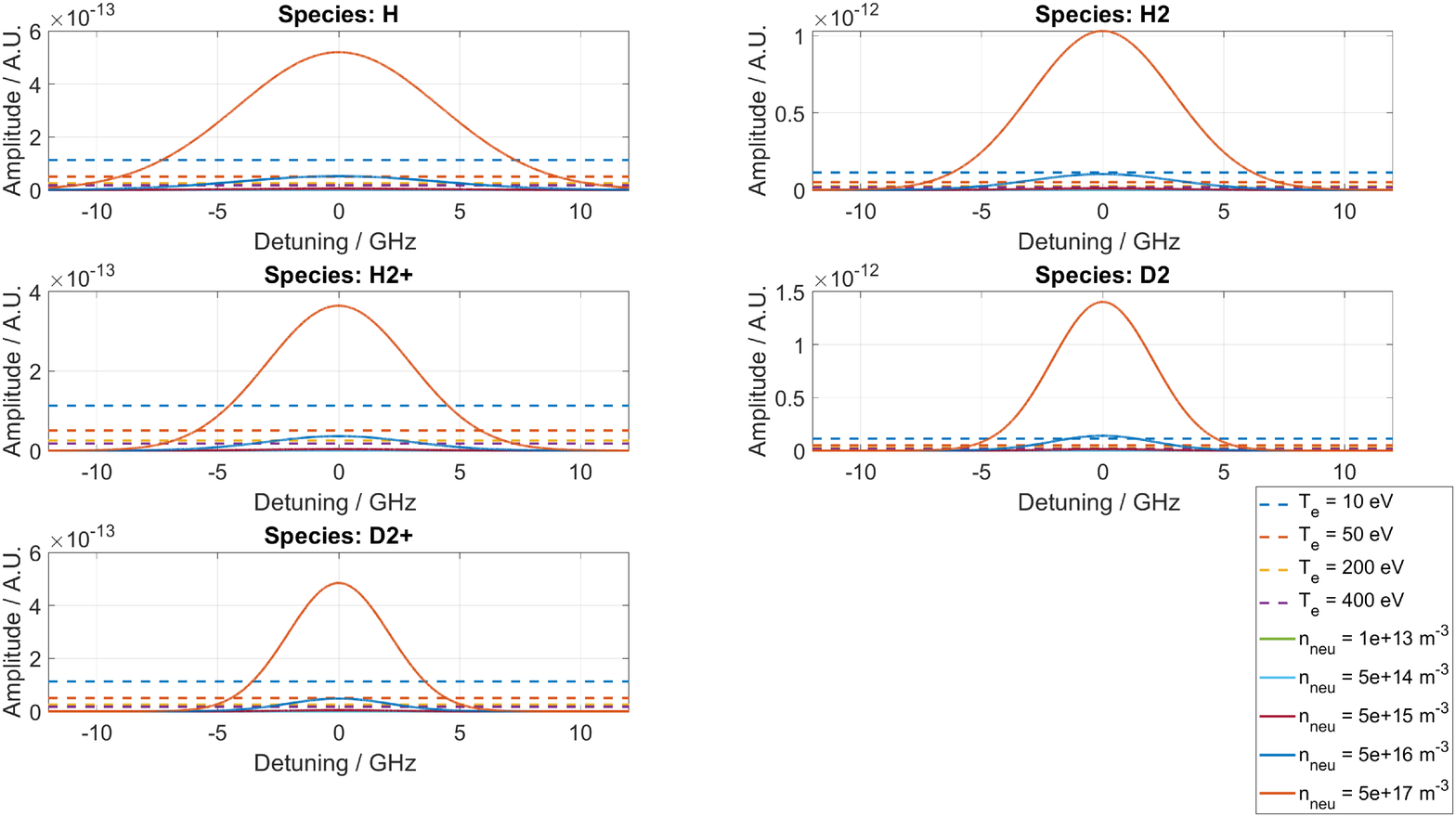}
	\caption{RS spectra (solid lines) for multiple species at neutral densities ranging from $ 1 \ \times 10^{13} \ \text{m}^{-3} $ to $ 5 \times 10^{17} \ \text{m}^{-3} $ in the presence of plasma backgrounds (dashed lines) at $ n_e = 1 \times 10^{17} \ \text{m}^{-3} $. Each spectra corresponds to the species listed above it.}
	\label{spectra_multispecies_1e17}
\end{figure}

\begin{figure}[H]
	\centering
	\includegraphics[scale=0.35]{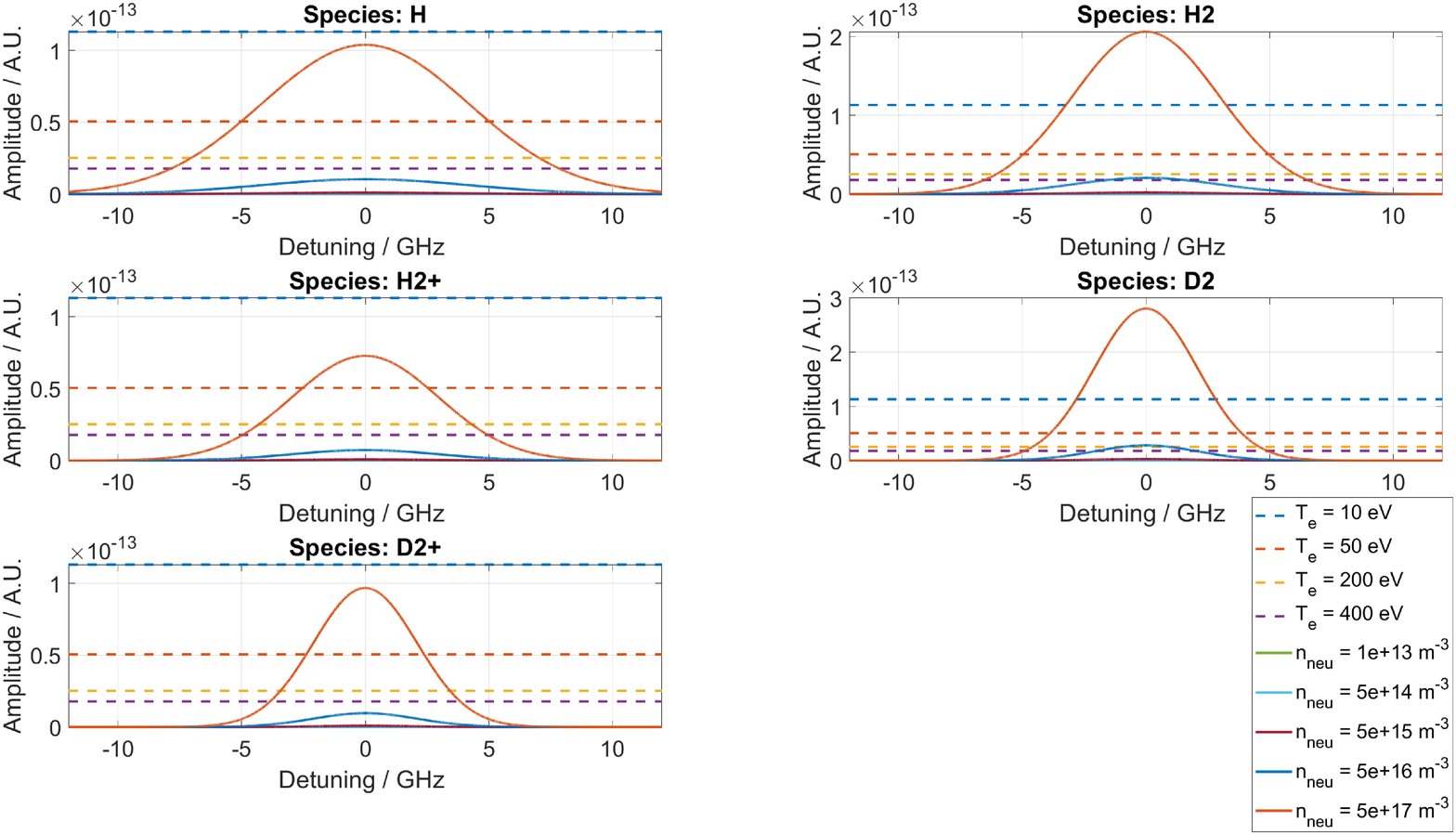}
	\caption{RS spectra (solid lines) for multiple species at neutral densities ranging from $ 1 \ \times 10^{13} \ \text{m}^{-3} $ to $ 5 \times 10^{17} \ \text{m}^{-3} $ in the presence of plasma backgrounds (dashed lines) at $ n_e = 5 \times 10^{17} \ \text{m}^{-3} $.}
	\label{spectra_multispecies__5e17}
\end{figure}

Figure \ref{spectra_multispecies_1e17}-\ref{spectra_multispecies__5e17} show the remaining profiles for the considered species at electron temperatures and neutral densities corresponding to $ R = $ 1.4-1.5 m up for $ n_e = 1\text{-}5 \times 10^{17} \ \text{m}^{-3} $. Note that for H and the molecular cations, the RS peak is unresolvable because their polarizabilities (and therefore their cross-sections) are lower relative to the other molecules', which results in a lower scaling factor according to Eqn. \ref{g_sc}.

The volumetric photon emission $ \Phi $ can also be calculated as follows:

\begin{align}
    \Phi = \frac{P}{h \nu \cdot V}
\end{align}

where the units are photons per unit volume per unit time. Figure 6 shows $ \Phi $ for the experimental configuration $ E_\text{L} = 0.5 $ J, $ \tau_P = 10 $ ns, $ \lambda = 532 $ nm, at right angles for both the azimuthal and polar angle, a distance to the photodetector 1 m, and radius of solid angle 1 cm, over neutral densities from $ 10^{14} $ to $ 10^{17} \ \text{m}^{-3}$. The power is integrated over the spectral domain shown in the previous figures.

\begin{figure}[H]
	\centering
	\includegraphics[scale=0.5]{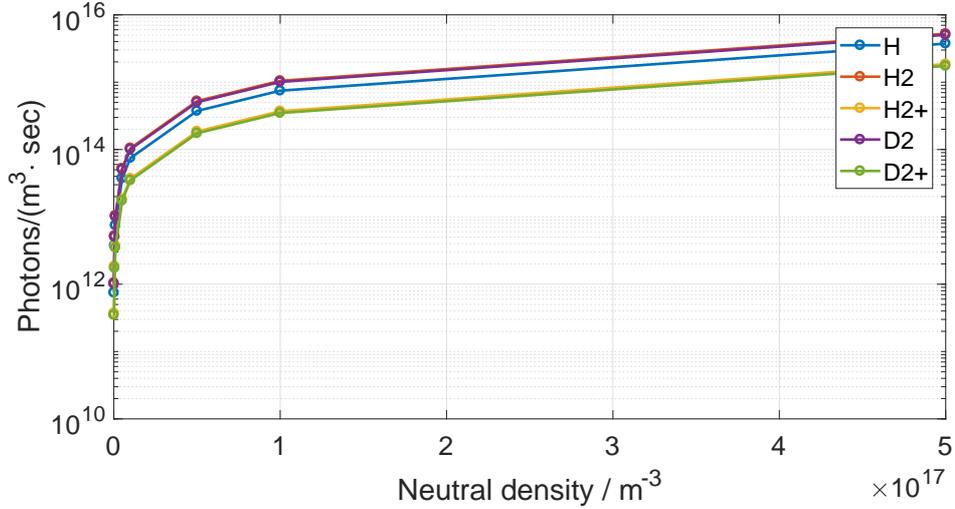}
	\caption{Volumetric photon emission of multiple species ranging from neutral densities $ 1 \times 10^{13} \ \text{m}^{-3} $ to $ 5 \times 10^{17} \ \text{m}^{-3} $ for $ n_e = 1 \times 10^{17} \ \text{m}^{-3} $.}
	\label{photonflux_multispecies}
\end{figure}

For a beam waist of $ w_\text{L} = 1 $ mm, this gives an area $ A = (\pi w_\text{L}^2) / 4 \approx 7.854 \times 10^{-7} \ \text{m}^2 $ and a Rayleigh range $ z_R = (\pi w_\text{L}^2) / \lambda \approx 0.059 \ \text{m} $, and thus an interrogation volume of $ V = A \cdot 2z_R \approx 9.276 \times 10^{-6} \ \text{m}^{3}$. For the neutral density on Figure \ref{spectra_multispecies_1e17} (approximately $ \approx 10^{17} \ \text{m}^{-3} $), this gives a range of about $ 1.616 \times 10^{10} $ to $ 4.859 \times 10^{10} $ photons per second; for $ \tau_p = 8 $ ns, this is approximately $ 3.89 \times 10^2 $ photons per pulse. For neutral densities below it, it becomes extremely challenging to generate photons. In comparison, an electron density of about $ 1 \times 10^{17} \ \text{m}^{-3} $ leads to a photon emission of $ 5.329 \times 10^{13} $ photons per second, or $ 4.263 \times 10^5 $ photons per pulse for the same ns pulse, which is three orders higher than for the largest neutral density in Figure \ref{photonflux_multispecies}. \\

\begin{figure}[H]
	\centering
	\includegraphics[scale=0.4]{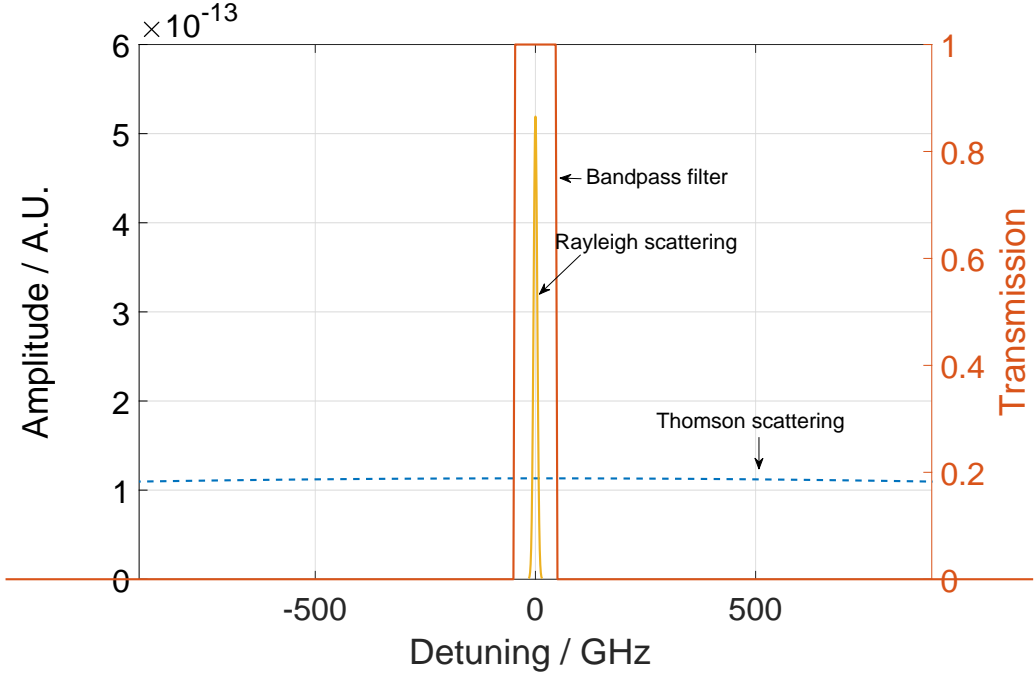}
	\caption{Representation of bandpass cutoff (orange line) of TS (dashed line), leaving primarily the RS (yellow line). The fraction of the scattered power is the integrated signal inside the bandpass region.}
	\label{BPfilter}
\end{figure}

When integrating across the entire spectral domain, it would seem that $ \Phi $ from the TS is greater than that of the RS, a threshold below which would make the RS signal undetectable. The idea would then be to reduce the bandwidth and focus over a few tens of GHz according to the linewidth of the RS. Consider a bandpass filter narrow enough to eliminate the transmission across the entire spectrum except for the domain $ \nu = [ -1, 1 ] $ THz as shown in Figure \ref{BPfilter}.
 
\begin{figure}[H]
	\centering
	\includegraphics[scale=0.5]{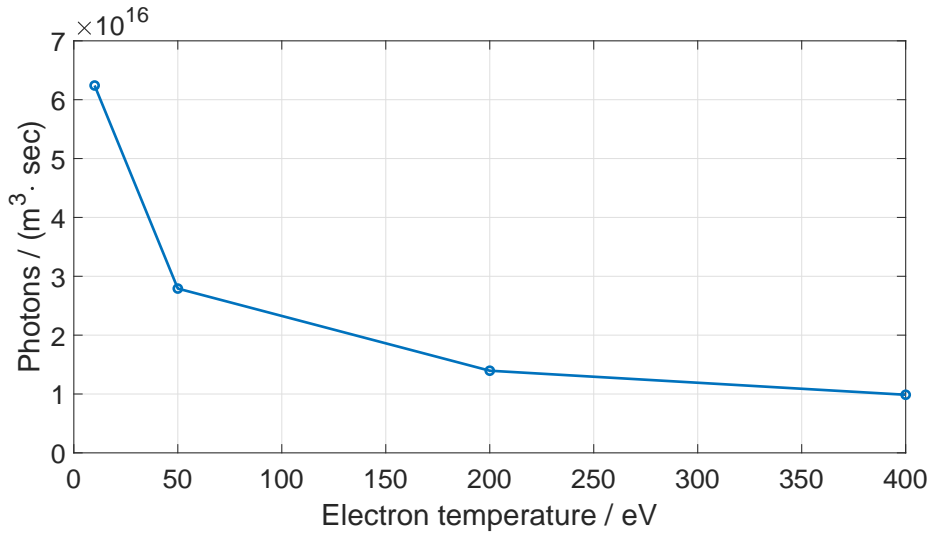}
	\caption{Filtered-out TS powers based on the bandpass filter region ($ \Delta \nu = \pm 1 $ THz) in Figure \ref{BPfilter}. Electron density is $ n_e = 1 \times 10^{17} \text{m}^{-3} $.}
	\label{photonflux_electron}
\end{figure}

By integrating over this region only, for some number densities we are able to retrieve a greater fraction of the photons coming from the RS than the TS scattering, and thus a detectable RS signal to the photodetector. The result is shown in Figure \ref{photonflux_electron}.

At electron temperatures above 50 eV, the filtered-out TS photon emission is reduced to the same order of magnitude as $  n_\text{neu} \approx 5 \times 10^{17}  m^{-3} $ and lower in magnitude for the cases of \ce{H}, \ce{H_2}, and \ce{D_2}. If the bandpass domain could be reduced by one order of magnitude further (e.g. $ \nu = [-100,100] $ GHz, or approximately $ \lambda = [-100,100] $ pm) then the TS photon emission can be further reduced by one order of magnitude. It will be a challenge as this requirement pushes the limits of bandpass filter technology. \\


\section{Preliminary Studies: Small Angle Slot Divertor}

The concept of a divertor to efficiently dissipate heat from tokamaks is critical for high performance (H-mode) operation. The divertor is situated along the magnetic field lines of the tokamak and ``divert" the plasma away from other walls which could be eroded by its interaction with the high temperature plasmas. To achieve near-zero erosion at the divertor surface, the divertor plasma temperature must be kept cool by diffusing gas at the build-up regions, or the slots, in the divertor. An accurate dtermination of the neutral profile in the divertor will provide key pararemeters for an accurate modeling of the radiation in future fusion devices.\\

\begin{figure}[H]
	\centering
	\includegraphics[scale=1]{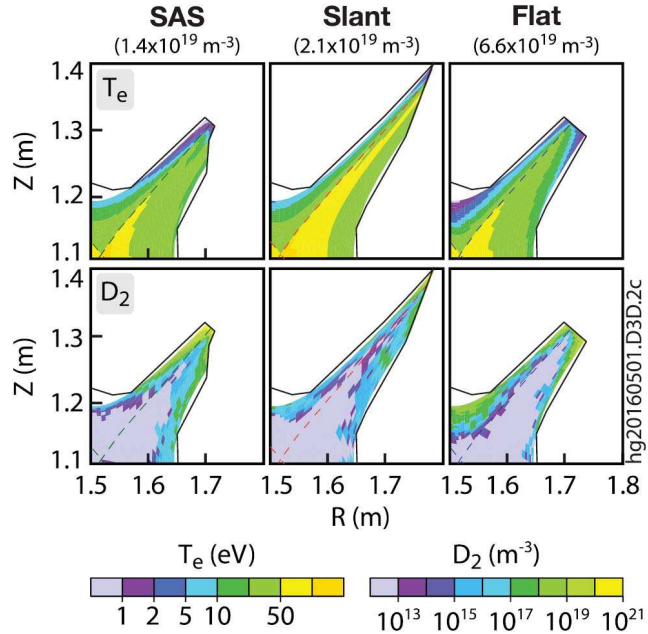}
	\caption{$ T_e $ and \ce{D_2} distributions in the SAS, slant, and flat-end divertors. Figure is obtained from \cite{guo_small_2017}.}
	\label{guo2017_SAS}
\end{figure}

The following preliminary results will focus on the neutral density conditions of the small angle slot (SAS) divertor proposed in Guo et al. \cite{guo_small_2017}. The most important part of the SAS divertor is its small angle at the outboard side end of the divertor slot, where electron temperatures must be kept sufficiently cool. These electron temperatures exhibit a close correlation to the \ce{D_2} neutral density distributions as shown in Figure \ref{guo2017_SAS}.

\begin{figure}[H]
	\centering
	\includegraphics[scale=0.5]{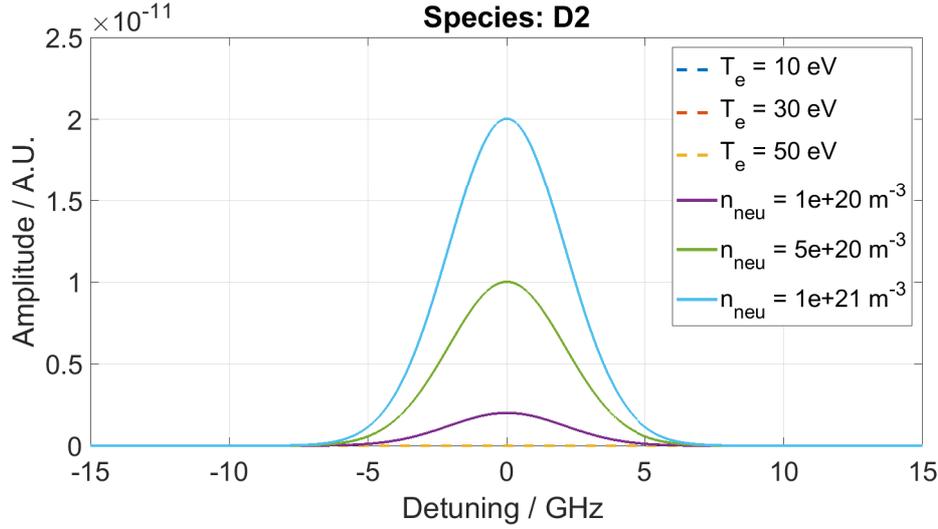}
	\caption{RS spectra (solid lines) for \ce{D_2} at neutral densities ranging from $ 1 \times 10^{20} \ \text{m}^3 $ to $ 1 \times 10^{21} \ \text{m}^3 $. The TS background spectra (dashed lines) are much lower than the RS spectra over the given spectral region.}
	\label{spectra_D2_divertor}
\end{figure}

Neutral densities are significantly higher than those in the outer midplane of the NSTX by up to six orders of magnitude. Such densities are practical for LRS and can be easily commensurate, assuming the majority of the TS background can be filtered out. Figure \ref{spectra_D2_divertor} shows the spectra of \ce{D_2} for $ \lambda = 532 $ nm, $ n_\text{e} = 1.4 \times 10^{19} \ \text{m}^{-3} $ and $ T_\text{e} = 10, 30, 50 $ eV. Spectra are all calibrated to the electron density. As shown, the \ce{D_2} spectra are well above the TS background.

\begin{figure}[H]
	\centering
	\includegraphics[scale=0.5]{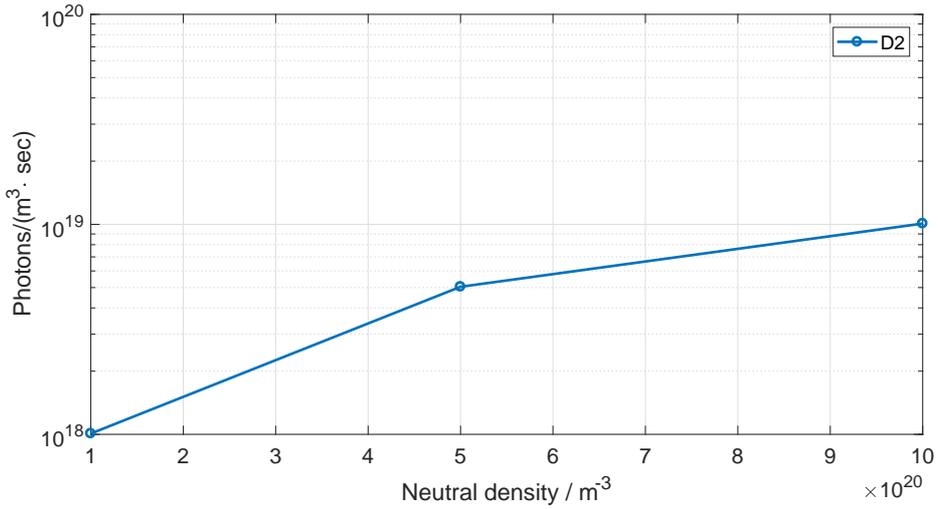}
	\caption{Volumetric photon emission for the neutral densities of Figure \ref{spectra_D2_divertor}.}
	\label{photonflux_D2_divertor}
\end{figure}

Figure \ref{photonflux_D2_divertor} shows the corresponding volumetric photon emission for $ E_\text{L} = 0.5 $ J, $ \tau_\text{p} = 8 $ ns, and $ w_\text{L} = 1 $ mm. A neutral density of $ n_\text{neu} = 1 \times 10^{21} \ \text{m}^{-3} $ gives $ 9.359 \times 10^{13} $ photons per second (assuming $ V = A \cdot b $ where $ b = 2 z_R $ is the confocal parameter), and thus \textit{$ 7.488 \times 10^5 $} photons per pulse. In comparison, an electron density of $ n_e = 1.4 \times 10^{19} $ gives $ 7.461 \times 10^{15} $ photons per second without filtering, $ 3.626 \times 10^{13} $ photons per second with a $ \Delta \nu = [-100,100] $ GHz bandpass filter. A neutral density measurement can therefore be achieved in the SAS divertor concept provided the TS can be filtered out at the wings. LRS results may also be mixed for the slant and flat slot divertors in Figure \ref{guo2017_SAS} since their electron densities and neutral densities each does not vary by more than one order of magnitude from the SAS configuration. \\


\section{Other Important Diagnostic Considerations}

From Eqn. \ref{DiffCrossSec}, we can significantly increase the RS power by decreasing the wavelength of the laser via the cross-section:

\begin{align}
     d_{\Omega} \sigma_{\text{RS}} \sim k^4 \sim \frac{1}{\lambda^4}
\end{align}

\begin{figure}[H]
	\centering
	\includegraphics[scale=.35]{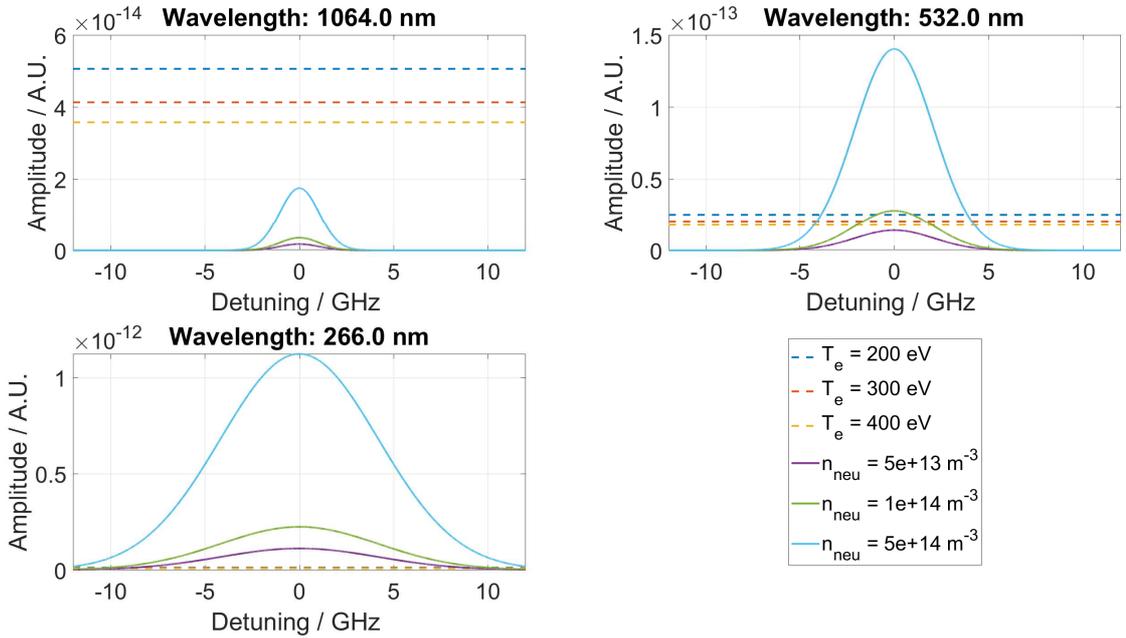}
	\caption{RS spectra (solid lines in TS background spectra (dashed lines) for \ce{D_2} and $ n_{\text{e}} = 10^{15} \ \text{m}^{-3} $.}
	\label{spectra_multicolor_D2}
\end{figure}

Consider the harmonic generation of an Nd:YAG laser: 1064 (fundamental), 532 (second harmonic), and 266 nm (fourth harmonic). The increase in RS power from 1064 to 532 nm is 16x, and from 532 to 266 is 16x, resulting in an overall 256x increase in total RS cross-section from the fundamental down to the fourth harmonic. However, the amount of energy per pulse is also decreased, typically by 70-80\% of the fundamental at the fourth harmonic. Figure \ref{spectra_multicolor_D2} compares spectra of \ce{D_2} for the three wavelengths at the conditions for R = 1.4 m in the NSTX with $ n_{\text{e}} = 10^{15} \ \text{m}^{-3} $. At $ \lambda = 1064 $ nm, the RBS spectra is completely buried underneath the TS spectra. At $ \lambda = 532 $ nm, the spectrum at $ n_{\text{neu}} = 1 \times 10^{14} \ \text{m}^{-3} $ barely peaks above the TS background, and at $ \lambda = 266 $ nm the RBS spectra is significantly above it. \\

\begin{figure}[H]
	\centering
	\begin{subfigure}{.5\textwidth}
	\includegraphics[scale=.34]{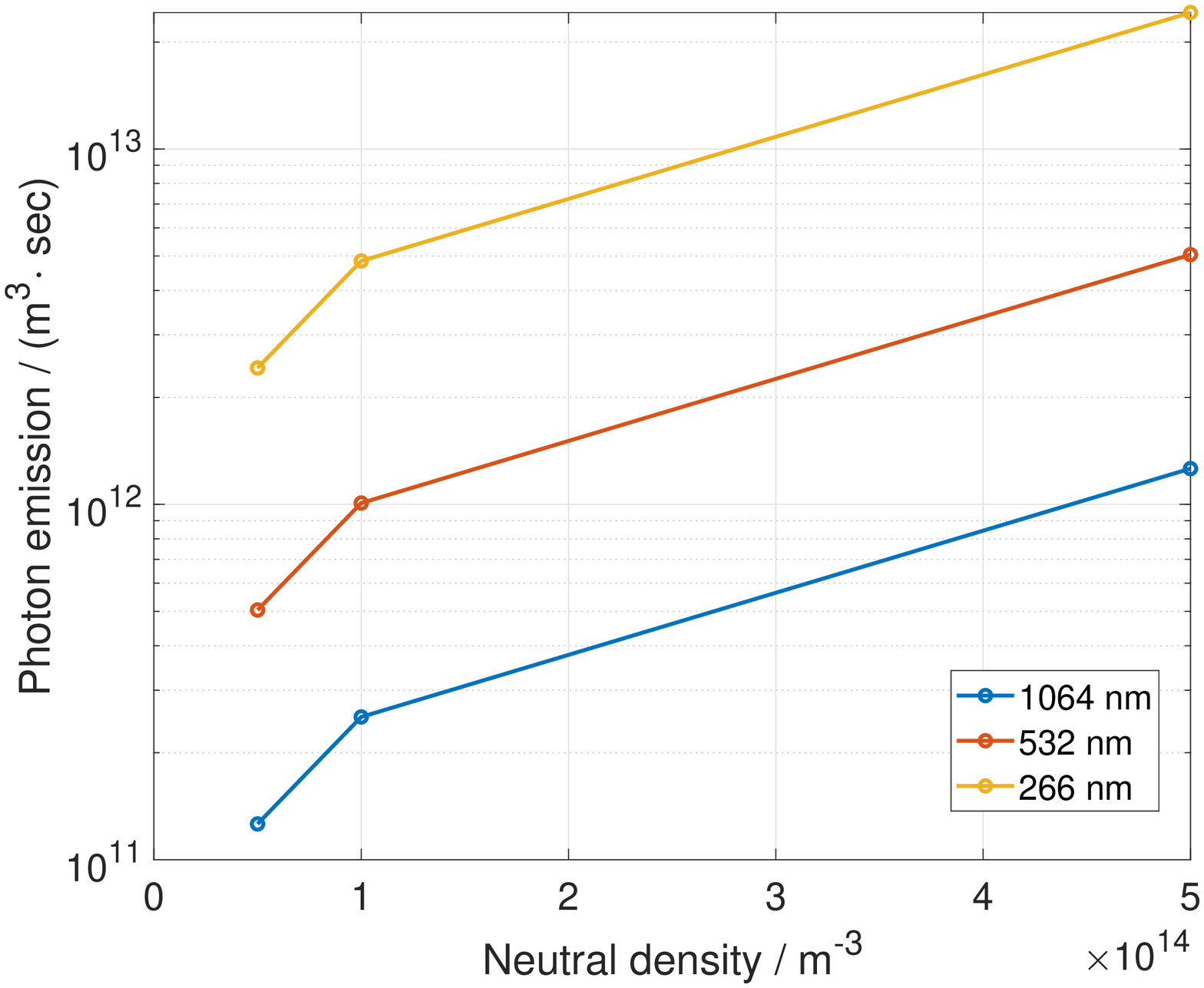}	
	\end{subfigure}%
	\begin{subfigure}{.5\textwidth}
	\includegraphics[scale=.34]{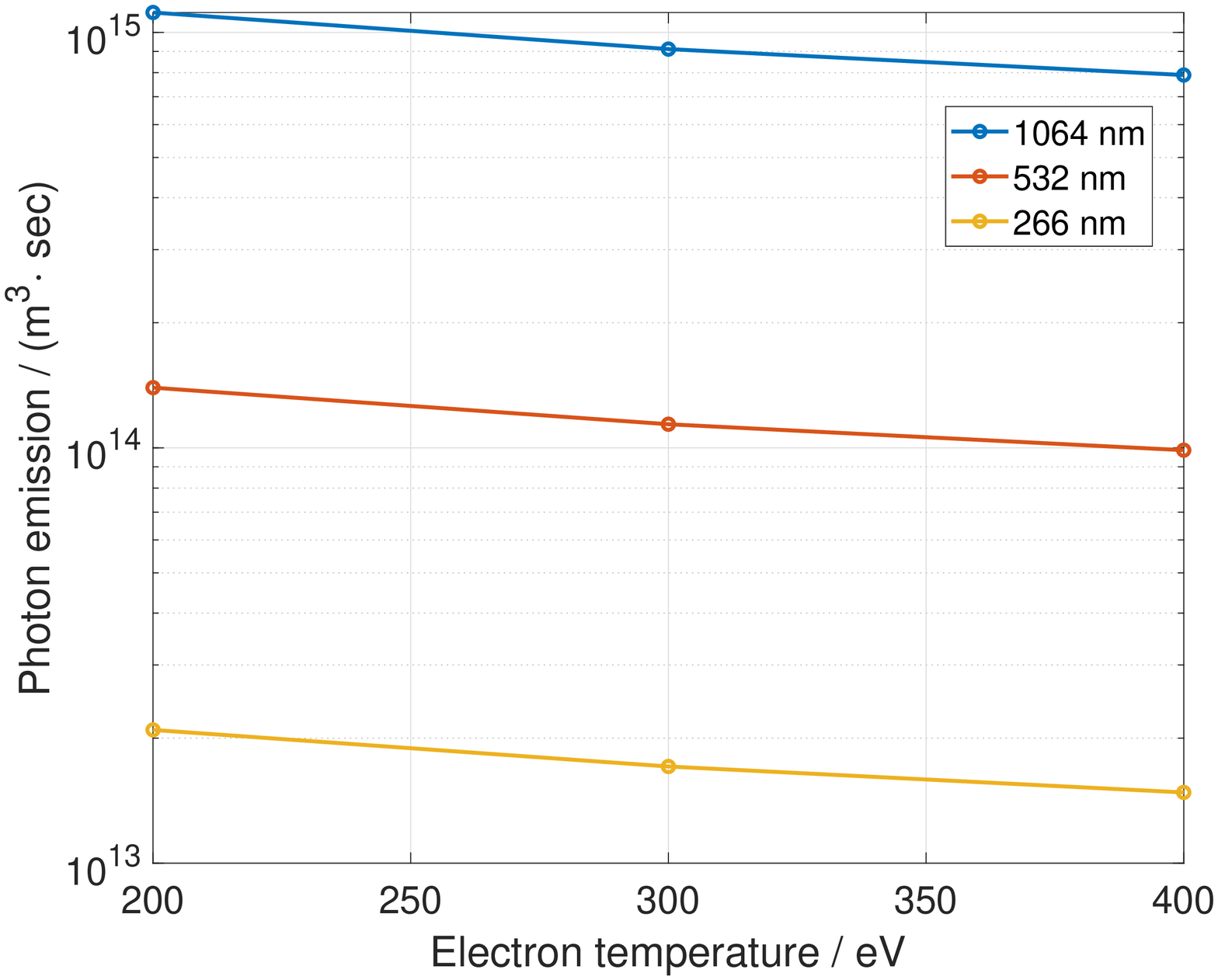}	
	\end{subfigure}
	\caption{Volumetric photon emission for \ce{D_s} (left) and electron (right) at multiple wavelengths of an Nd:YAG laser. Neutral density and electron temperature ranges mirror those of the NSTX midplane profiles (see Figure \ref{stotler}).}
	\label{photonflux_multicolor_D2}
\end{figure}

The photon volumetric emission $ \Phi $ for the neutrals corresponding to Figure \ref{spectra_multicolor_D2} and electron is shown in Figure \ref{photonflux_multicolor_D2}. The energy per pulse for $ \lambda = 1064, 532, 266 $ nm is $ E = 1, 0.5, 0.3 $ J, respectively. The width of the filter to suppress the TS is 100 GHz centered at the laser center. Note that $ \Phi_{\text{neu}} $ at $ \lambda = 266 $ nm is about two orders of magnitude greater than that of 1064 nm, while $ \Phi_\text{e} $ suffers because it is only dependent on the pulse energy and not the wavelength. Using the parameters $ \lambda = 266 $ nm, $ w_\text{L} = 1 \ \text{mm} $, and thus $ V = A \cdot b  \approx 1.855 \times 10^{-5} \ \text{m}^3 $, we obtain about $ 2.994 \times 10^{8} $ photons per second for $ n_\text{neu} = 5 \times 10^{14} \text{m}^{-3} $; for a pulse width of 8 ns, this corresponds to about 2.4 \textit{photons per pulse}. Calculations were also done for $ w_\text{L} = 10 $ mm which resulted in about 100 times more photons per pulse. A larger beam parameter is therefore more advantageous and desirable assuming the entire confocal length region can be captured, although it is still a very small number of photons to collect, making the NSTX midplane region a highly challenging if not impossible task to do LRS.

\begin{figure}[H]
	\centering
	\begin{subfigure}{.5\textwidth}
	\includegraphics[scale=.34]{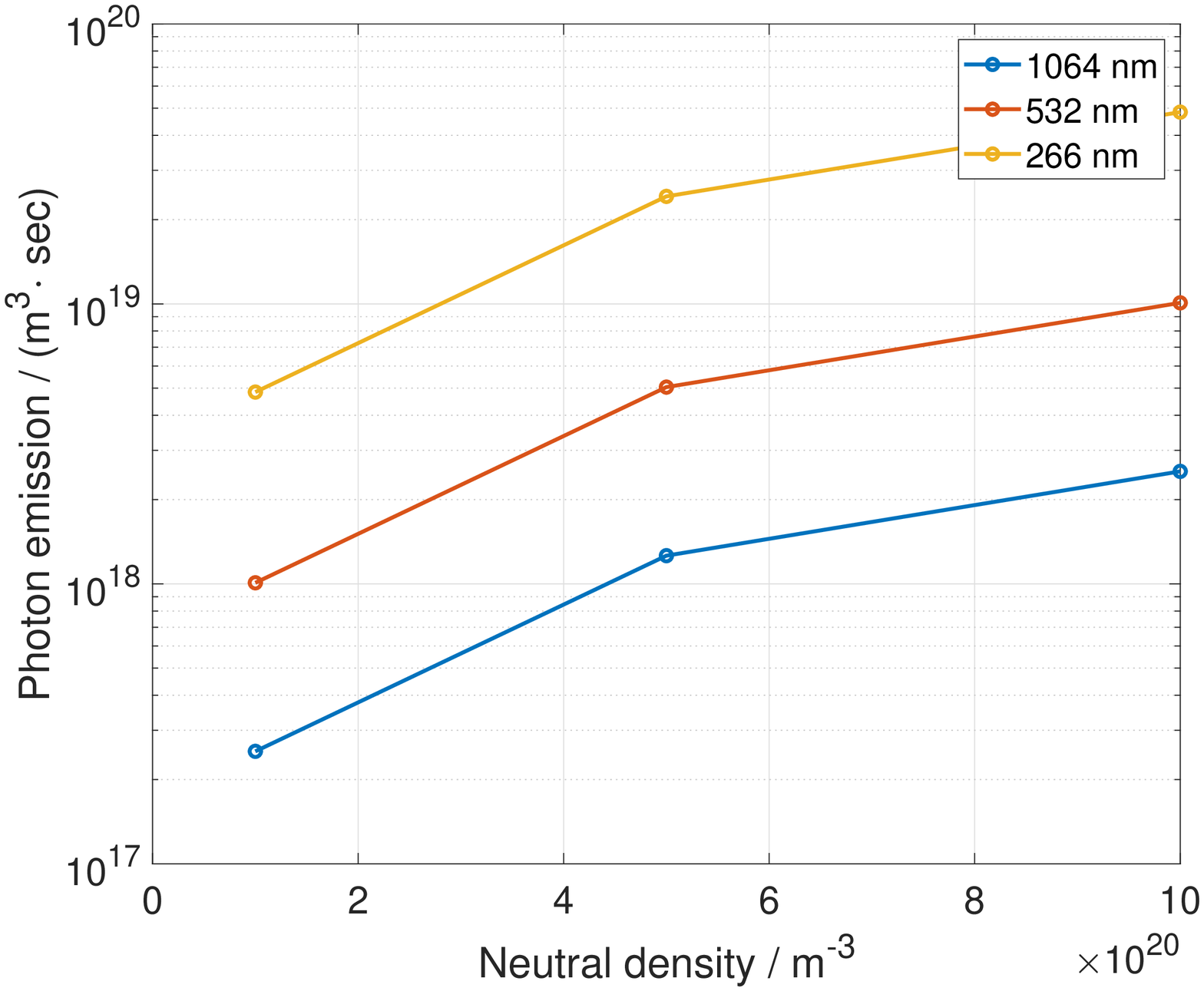}
	\end{subfigure}%
	\begin{subfigure}{.5\textwidth}
	\includegraphics[scale=.34]{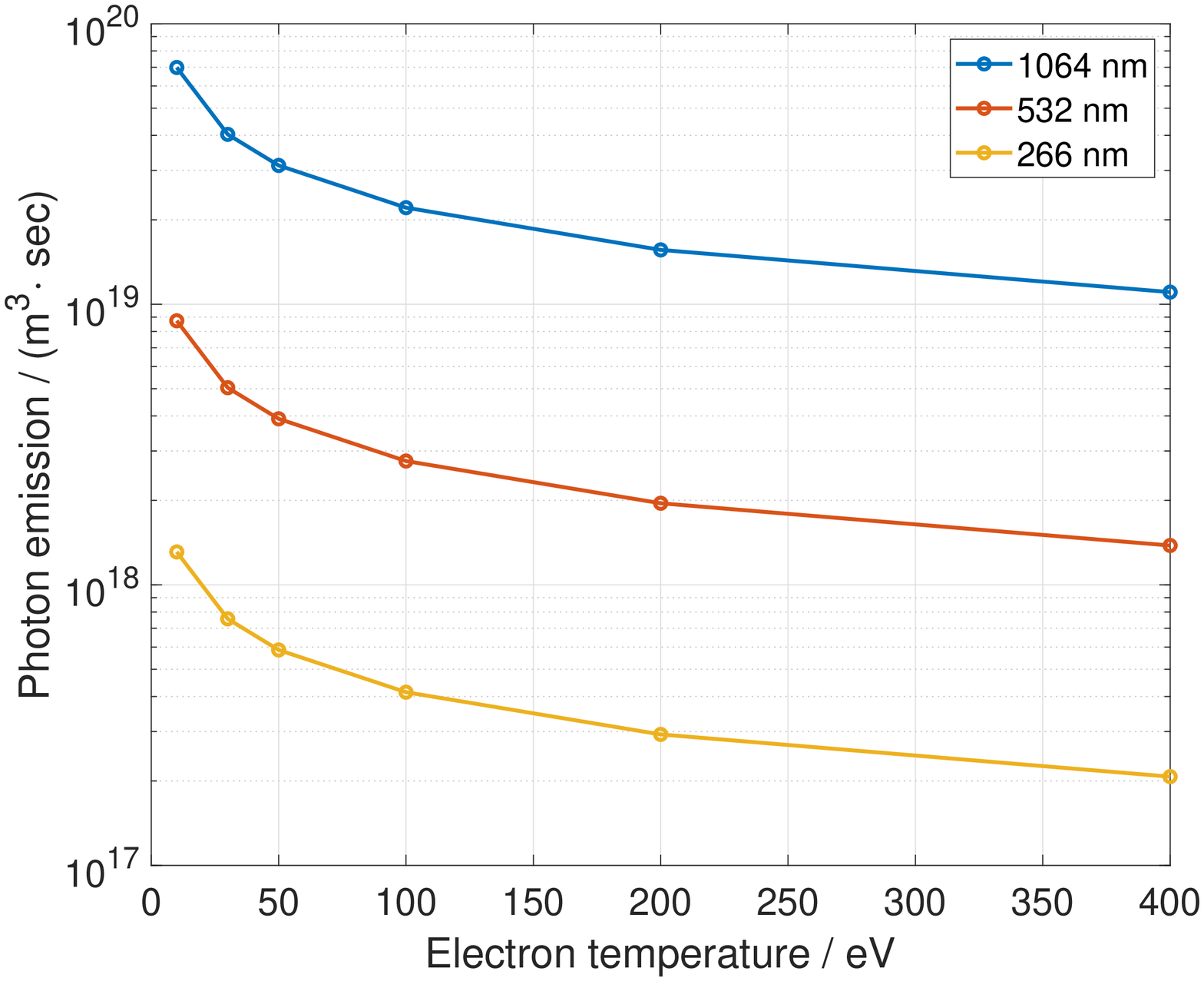}	
	\end{subfigure}
	\caption{Volumetric photon emission for \ce{D_s} (left) and electron (right) at multiple wavelengths of an Nd:YAG laser. Neutral density and electron temperature ranges mirror those of the SAS divertor (see Figure \ref{guo2017_SAS}).}
	\label{photonflux_multicolor_divertor}
\end{figure}

Similarly, $ \Phi $ is shown for divertor conditions in Figure \ref{photonflux_multicolor_divertor}. The neutral densities are higher by an order of magnitude of about 6-7, and thus a greater photon emission by approximately the same order of magnitude. This results in about \textit{$ 4.791 \times 10^6 $ photons per pulse} at the divertor. Even for interrogation lengths which are restricted due to detector size or collection optics, they would pose a decrease of at most one to two orders of magnitude.

Below is a table that summarizes the photons generated for the various neutral densities for \ce{D_2}. Note that the number of photons per pulse is directly proportional to the neutral density.

\begin{table}[H]
\centering
\begin{tabular}{ |c|l|l|l| }
\hline
Neutral density ($ \text{m}^{-3} $) & Photons per pulse \\ \hline
$ 10^{14} $ & 0.479 \\ \hline
$ 10^{17} $ &  $ 4.791 \times 10^4 $ \\ \hline
$ 10^{19} $ & $ 4.791 \times 10^6 $ \\ \hline
$ 10^{20} $ & $ 4.791 \times 10^7 $ \\ \hline
$ 10^{21} $ & $ 4.791 \times 10^8 $ \\ \hline
\end{tabular}
\caption{Summary of photons generated for various neutral number densities from \ce{D_2}. The conditions used are $ E_\text{L} = 0.3 $ mJ, $ \tau_p = 8 $ ns, $ w_\text{L} = 1 $ cm, $ \lambda = 266 $ nm. Length is taken as the confocal beam parameter.}
\end{table}

The main consideration for wavelength will be the laser source. An Nd:YAG pulsed laser serves as the ideal laser system for this detection scheme due to its high peak powers in the visible and robust, long-term performance. Furthermore, it can be seeded or injection-locked for greater spectral purity of the laser wavelength; this is necessary for filtered Rayleigh scattering, which can reduce the highly narrow and much typically much stronger (than the RS) background scattering at the central wavelength. Overall, these aspects have made the Nd:YAG laser the workhorse of many LRS diagnostics and experiments, evidenced by the literature on LRS and its variations. However, output of the fourth harmonic ($ \lambda_\text{L} = 266 $ nm) is a different consideration, and modern commercial Nd:YAG lasers provide only around 100-250 mJ of fourth harmonic generation (see Spectraphysics Quanta-Ray Pro series and Surelite II series). Fourth harmonic lasers with up to 400-500 mJ have also been demonstrated in the laboratory \cite{suzuki_0.43_2002,wang_high-power_2012}. Although the energy change in comparison to wavelength is less consequential considering it is only around one order of magnitude, it is important for ensuring there are enough photons to collect in the measurement.

\section{Two-color Scattering Approach}

Despite the reduction in laser energy/intensity due to nonlinear generation, it is clear that a multi-color approach towards shorter wavelengths offers an advantage for LRS in plasma environments, both for enhancing the LRS detection signal and distinguishing between LRS and LTS, if there is ever a need to measure both densities. The way to do this would be via a \textit{two-color scattering} approach, which has been done for a laser spark shockwave in \cite{limbach_characterization_2015}, and is described as follows. \\

In some cases, it is not possible to determine from the measurement alone whether the photodetector signal primarily comes from LRS or LTS. However, as noted previously, the TS cross-section is independent of $ \lambda_\text{L} $ whereas the RS cross-section has a $ \lambda_\text{L}^{-4} $ dependence. Comparison of two different wavelengths provides a way to discriminate between the two, providing there are sufficient photons for scattering from each. This is known as two-color scattering. The total signal for one color is:

\begin{align}
    S(\lambda) = \frac{ n_\text{neu} \sigma_\text{RS}(\lambda) + n_\text{e} \sigma_\text{TS} }{ n_0 \sigma_\text{RS}(\lambda) } \label{twocolor}
\end{align}

All quantities are known except for $ n_\text{neu}, n_\text{e} $. Two colors give two equations with two unknowns, thereby providing a determinate system. The denominator of Eqn. \ref{twocolor} is the reference signal, typically in atmospheric conditions.

\section{Summary and Discussion}

{\textbf{Summary:}}
Using laser Rayleigh scattering for neutral measurements in tokamaks has long been known to be challenging. Simulations show that the Rayleigh peak can be resolved under relatively comparable electron densities from $ 10^{15} $ to $ 10^{19} \ \text{m}^{-3} $ and low to moderate electron temperatures (consistent with tokamak divertor temperatures). Calculations for NSTX midplane profiles ($ n_\text{neu} = 10^{13} $-$ 10^{17} \ \text{m}^{-3} $) imply photon generation per pulse to be virtually zero at $ \lambda_\text{L} = 532 $ nm and nonzero yet still ``counting photons" at 266 nm (fourth harmonic of Nd:YAG laser). Moreover, for the same calculations at the divertor region ($ n_\text{neu} = 10^{20} $-$ 10^{21} \ \text{m}^{-3} $), we obtain approximately $ 10^6 $ photons per pulse. Going down to the ultraviolet regime is crucial for generating more RS photons since the configuration at standard frequency-doubled Nd:YAG wavelengths is starved at the neutral density ranges of up to $ 10^{21} \ \text{m}^{-3} $. Narrow bandpass filters that can limit passing regions of up to $ [-1,1] $ THz will need to be implemented, and possibly even narrower (hundreds of GHz, or hundreds of pm), in order to filter out at least one to two orders of magnitude of Thomson scattering. Lastly, provided the RS power is significant enough, a two-color scattering approach can be implemented to separate the RS from the TS. Two-color scattering is critical to this diagnostic effort of resolving neutral densities directly for the first time in tokamaks. \\

{\textbf{Discussion:}}
While we show that in the midplane, the Rayleigh scattering photon count is too low to resolve the neutral densities, it is conceivable to utilize this approach to get to the neutral penetration into the edge/pedestal of tokamaks. More specifically, it is well-known that the neutral opacity of ITER is still unknown. This diagnostic could enable the study of the pedestal density formation under gas fuelling (to be in the field of view of diagnostic resulting into a doubly active diagnostics) in present day tokamaks, which could be used to unravel the degree of neutral opacity for future devices such as ITER. \\
\\

\textbf{Acknowlegment:} The authors are grateful to Prof. R.B. Miles for his interest in the work and valuable comments. This work was funded under the fast track 2019 LDRD project ``Feasibility Study of a Laser-Based Approach for Diagnosing Deuterium Neutrals in the Edge of Fusion Devices."


\bibliographystyle{plain}
\bibliography{main}

\end{document}